\documentstyle[12pt,cite]{article}

\begin{document}

\begin{center}

{\bf The Mikheyev-Smirnov-Wolfenstein Effect in Moving Matter}

{\bf A.Grigoriev, A.Lobanov,
A.Studenikin\footnote{\normalsize E-mail: 
studenik@srd.sinp.msu.ru, studenik@srdlan.npi.msu.su}}


{\bf Department of Theoretical Physics, Moscow State University,
119899 Moscow, Russia}

{\it Abstract}
\end{center}
{\it The Lorentz invariant formalism for description 
of neutrino flavour oscillations in moving matter is developed. It is shown that 
the matter motion 
with relativistic speed can sufficiently change the matter term in the neutrino 
effective potential.
In the case of matter motion parallel to neutrino 
propagation, matter effects in neutrino flavour oscillations are suppressed. 
In the case of relativistic motion of matter 
in the opposite direction in respect to neutrino propagation, sufficient 
increase of effects of matter in neutrino oscillations is predicted.
The effective neutrino flavour oscillation probability, effective mixing 
angle and oscillation length, as well as the MSW effect
resonance condition for the case of moving matter are
derived.}


The problem of neutrino propagation in matter, starting from the paper 
of Wolfenstein \cite {Wol78} 
is still under intensive study. The most famous 
result in this field has been  received when Mikheev and Smirnov \cite {MikSmi85}
showed that the probability of conversion between different neutrino flavour 
states can be increased (the MSW effect) due to neutrino interaction with 
particles of 
matter if the resonance condition is fulfilled even in the case of very small 
vacuum mixing angle. Interaction of neutrino with background 
matter or external electromagnetic fields can also produce modification of 
neutrino characteristics. For example, under the influence of external environment 
massless
neutrino can acquire an effective mass \cite {Wel82} and also 
non-trivial electromagnetic properties 
\cite {DOlNiePal89} 
like non-zero magnetic moment.  It is commonly believed that neutrino, propagating through matter 
and being under the influence of electromagnetic fields might exhibit their properties that under 
such influence become accessible for experimental observation. That is why we
consider below new phenomena in neutrino oscillations that could be 
produced by motion of matter through which neutrino is propagating.

In a series of papers \cite {ELSt99, ELStpl00, ELSt00} we have 
recently developed 
the Lorentz invariant formalism for neutrino motion in
non-moving and isotropic matter
under the influence of an arbitrary configuration of electromagnetic fields.
The effective Hamiltonian for the neutrino spin evolution, which accounts
not only for the transversal to the neutrino momentum components of
electromagnetic field but also for the longitudinal components have been derived. With
the using of the proposed Hamiltonian it becomes  possible to consider
neutrino spin precession in an arbitrary configuration of
electromagnetic fields including those that contain strong
longitudinal components. We have considered the new types
of resonances in the neutrino spin precession
$\nu_{L}\leftrightarrow\nu_{R}$ that could appear when neutrinos propagate
in matter under the influence of different electromagnetic field
configurations (see also \cite {DvoStuYa01}). Within the proposed approach the parametric 
resonance of neutrino oscillations in electromagnetic wave field with periodically
varying time-dependent amplitude and the periodic "castle wall" magnetic field has been also 
studied \cite {DSthp0102099}.

In the studies \cite {ELSt99,ELStpl00, ELSt00} of the neutrino spin evolution
we have focused mainly on description of influence of different
electromagnetic fields, while modelling the matter we confined
ourselves to the most simple case of non-moving and unpolarized
matter.  In  \cite {LobStuph01} ( see also \cite {DELSt01}) we have further generalized our
approach and developed the covariant description of neutrino oscillations
in moving and polarized matter under the influence of electromagnetic fields. 
The important point is that this approach is valid for accounting of matter motion and polarization 
for arbitrary speed of matter. It has been shown for the first time that in the case of 
relativistic motion of matter, the value of effects of matter in neutrino spin (or spin flavour)
oscillations sufficiently 
depends on the direction of matter motion in respect to neutrino propagation and also on 
the values of matter and neutrino speeds.

It should be also noted here that effects of matter polarization in 
neutrino oscillations were considered previously in several papers
( see, for example, \cite {NSSV97,BGN99} and references therein). 
However, the used in refs. \cite {NSSV97,BGN99} procedure of
accounting for the matter polarization effect does not enable one to study 
the case of matter motion with total relativistic speed. 
Within our approach we can reproduce corresponding results of 
\cite {NSSV97,BGN99} in the case of matter which is slowly moving or 
is at rest.

As the main result of our studies \cite {DELSt01, LobStuph01} 
we have 
derived the equation for the neutrino
spin evolution in electromagnetic field $F_{\mu \nu}$ in moving and 
polarized matter. Let us now consider neutrino propagation 
in the relativistic flux of 
electrons in presence of external magnetic field. In the case the standard 
model of interaction supplied with
$SU(2)$-singlet right-handed neutrino $\nu_R$ for the 
evolution equation of the three-di\-men\-sio\-nal neutrino spin vector 
$\vec S$ accounting for the direct neutrino 
interaction with the magnetic field
and moving matter (the flux of electrons) it is possible to get 
\cite {LobStuph01}
\begin{equation}
{d\vec S \over dt}={2\mu \over \gamma} \Big[
{\vec S \times ({\vec B_0}+\vec M_0)} \Big], 
\end{equation}
where the derivative in the left-hand
side of eq.(1) is taken with respect to time $t$ in the
laboratory frame, whereas the value $\vec B_0$ is
the magnetic field in the neutrino rest frame given in terms
of the transversal in respect to the
neutrino motion ($\vec B_{\perp}$) and longitudinal ($\vec B_{\parallel}$) 
field components in the laboratory frame,

\begin{equation}
\begin{array}{c}
\vec B_0=\gamma\big(\vec B_{\perp}
+{1 \over \gamma} \vec B_{\parallel} \big).
\label{2}
\end{array}
\end{equation}

The influence of matter on the neutrino spin evolution in (1) is given by
the vector $\vec M_0$, for which 
in the case of slowly moving ($v_e\ll1$) and unpolarized matter we have got
\begin{equation}
{\vec M}_0=n_e\gamma{\vec\beta}
\rho^{(1)},
\label{3}
\end{equation}
where $n_{e}$ is the electron number density, 
$\vec \beta$ is the speed of neutrino 
($\gamma=(1-\beta)^{-1/2}$) and
\begin{equation}
\rho^{(1)}_e={G_F \over
{2\mu \sqrt2}}(1+4\sin^2 \theta _W).
\label{4}
\end{equation}
Formula (\ref{3}) reproduces the matter term for the case of neutrino spin oscillations which is similar to
the Wolfenstein term \cite {Wol78} in the neutrino effective potential in the case of flavour oscillations.
 In the opposite case of relativistic flux of electrons,
$v_e\sim 1$, we have found,
\begin{equation}
{\vec M}_0=n_e\gamma{\vec\beta}
\rho^{(1)}
\Big(
1-\vec\beta{\vec v}_e
\Big).
\label{5}
\end{equation}
Here for simplicity we neglect effects of matter polarization 
(contrary to what has been done in \cite 
{LobStuph01}).
If we
introduce the invariant electron number density, 
\begin{equation}
\begin{array}{c}
n_0=n_e\sqrt{1-v^2_e},
\end{array}
\label{6}
\end{equation}
then it follows,
\begin{equation}
\begin{array}{c}
{\vec M}_0=n_o\gamma{\vec\beta}
{(1-\vec\beta{\vec v}_e)\over\sqrt{1-v^2_e}}
\rho^{(1)}.
\label{7} 
\end{array}
\end{equation}
Thus, in the case of the parallel motion of neutrinos
and electrons of the flux, the matter effect contribution to the neutrino
spin evolution equation (1) is suppressed. In the case of neutrino 
and matter relativistic motion ($\beta$ and $v_e \sim 1$) 
in opposite directions, the matter term $\vec M_0$ gets its maximum 
value
\begin{equation}
\vec M_{0}^{max}=2n_e\gamma{\vec\beta}
\rho^{(1)},
\end{equation}
which is equal to the matter term 
derived for the case of slowly moving $(v_e \ll 1)$ matter times 
a factor ${2 \over \sqrt{1-v_{e}^2}}$. Therefore, as it has been predicted 
in \cite {LobStuph01} 
for the neutrino spin and spin flavour oscillations in matter and electromagnetic fields,  there can be 
sufficient
increase of matter effects in neutrino oscillations for neutrino  
propagating against the relativistic flux of matter. On the contrary, the matter effect can be  "eaten" by the 
relativistic motion of matter in the case when  matter is moving with relativistic total speed along the direction of 
neutrino propagation.

As one of our concluding remarks in \cite {LobStuph01} we have mentioned that the similar matter 
term suppression (or increase) effect might exist in neutrino flavour oscillation without change of 
helisity for the case of matter moving with relativistic speed. In this paper we prove 
this statement . In the frame of the Lorentz invariant approach to description of the 
neutrino flavour oscillations in matter 
we get the effective neutrino evolution Hamiltonian which can be used for the case of  
matter motion with arbitrary ( and also relativistic) total velocities. 
We show that the matter motion could lead to sufficient 
change in the neutrino flavour oscillation probabilities and other neutrino oscillation parameters 
(like the effective mixing angle and oscillation length) if matter is moving with relativistic speed.
In particular, in the case of matter moving parallel to the neutrino 
propagation, the Wolfenstein matter term in the neutrino effective potential is suppressed.  Contrary 
to this suppression effect, in the case of relativistic motion of matter in the opposite direction in 
respect to neutrino propagation, sufficient increase of effects of matter in neutrino flavour 
oscillations is predicted. We also argue that effects of 
matter motion have to be accounted for in 
the resonance conditions for the neutrino flavour oscillations.

Our goal is to investigate neutrino oscillations characteristics in 
the case of relativistic motion of matter which can be composed of different background fermions,
$f=e,n,p,\mu,\tau$ etc. In the general case each of the matter components is characterized by its own
distribution density function, the number density $n_{f}$ and the speed of the reference frame in 
which the mean momentum of the fermions $f$ is zero. The fermion $f$ current is determined as
\begin{equation}
j^{\mu}_{f}=(n_f, n_f \vec v_f).
\label{curr}
\end{equation}
If a component of matter $f$ is slowly moving or at rest in the laboratory frame, $\vec v_f \approx 0$, 
the fermion current equals
\begin{equation} 
j^{\mu}_{f}=(n_f, 0,0,0).
\label{curr_0}
\end{equation}

Let us suppose that at least one of the matter components $f$ is moving as a whole with relativistic speed,
$v_f \approx 1$. For simplicity, let us consider 
neutrino two-flavour oscillations, e.g. $\nu_{e} \leftrightarrow \nu_{\mu}$, in matter composed of  
the only one component, electrons ($f=e$), moving with relativistic total speed. 
Generalization for 
the case of the other types of neutrino conversions and different compositions and types of motion of 
matter is straight forward. 

The matter effect in neutrino oscillations occurs as the result of the elastic forward scattering of 
neutrinos off  the background fermions. In our case the difference $\Delta V$ between 
the mean potentials $V_{e}$ and 
$V_{\mu}$ for  the two flavour neutrinos is produced by the charged current interaction of  the electron 
neutrino with the background electrons. Note that the neutral current interaction is affective in oscillations 
between the active and sterile neutrinos. We do not account for effects of matter polarization which vanishes
if there is no preferred  spin orientation of the electrons. These effects were discussed in details for 
neutrino flavour oscillations in non-moving matter  \cite {NSSV97} and for neutrino spin oscillations 
in moving matter \cite {LobStuph01}.  We also neglect the momentum dependence of 
the charged vector boson propagator. Then the corresponding neutrino effective Lagrangian 
can be written in the following form (see \cite {LobStuph01} and also \cite {Pal92})
\begin{equation}
L_{eff}=-\sqrt {2} G_{F} j^{\mu}\Big(\bar \nu_{e_L}\gamma_{\mu} \nu_{e_L}\Big).
\label{L_eff}
\end{equation}
This additional term in the Lagrangian modifies the Dirac equation: 
\begin{equation}
\gamma_0 E-\vec \gamma \vec p -m=U\gamma_0 -U\vec v_{f}\vec \gamma, \ \ \ \ \ \ \ \ \ \ \ \ 
U=\sqrt {2} G_{F} n_{e}.
\label{Dir}
\end{equation}
Rearranging the terms we get the neutrino dispersion relation in matter in the following form,
\begin{equation}
E=\sqrt {(\vec p-U\vec v_{f})^2+m^2}+U.
\label{disp}
\end{equation}
The further simplification is found to occur in the limit of weak potential $Uv_{f} \ll \sqrt{\vec p ^{2}+m^2}$, so we 
get for the effective energy of the electron neutrino in the moving matter
\begin{equation}
E_{\nu_e} =\sqrt{\vec p ^{2}+m^2}+U(1-\vec \beta \vec v_{f})+O\Big({1 \over \gamma}\Big).
 \label{energy}
\end{equation}

The effective electron neutrino energy in addition to the vacuum energy $p_0=\sqrt{\vec p ^{2}+m^2}$ contains 
the Wolfenstein matter term which is proportional to the electron number density $n_e$. Note that there is no matter 
term in the muon neutrino effective energy in our case. The important new phenomenon, as it can be seen from 
(14), is the matter term dependence on the speeds of neutrino $\vec \beta$ and matter $\vec v_f$, and also on 
the angle between these two vectors. In the limit of slowly moving matter or matter at rest, $v_f \ll 1$, we get the usual expression for the matter term which does not depend 
on the speed of matter. 

Now following the usual procedure for the two flavour neutrinos, $\nu_e$ and $\nu_\mu$, with mixing 
in the high-energy limit it is 
possible to get the adiabatic solution for the neutrino oscillation problem. Thus, the probability 
of neutrino conversion $\nu_e \rightarrow \nu_\mu$ can be written in the form
\begin{equation}
P_{\nu_e \rightarrow \nu_\mu}(x)=\sin^{2} 2\theta_{eff} \sin^{2} {\pi x \over L_{eff}},
\label{ver}
\end{equation}
where the effective mixing angle, $\theta_{eff}$, and the effective oscillation length, $ L_{eff}$,
in the moving matter are given by
\begin{equation}
\sin^{2} 2\theta_{eff}={\Delta^{2}\sin^{2} 2\theta \over 
{\Big(\Delta \cos^2\theta - U(1-\vec \beta \vec v_{f})\Big)^2+ \Delta^{2}\sin^{2} 2\theta}},  
\label{th}
\end{equation}
\begin{equation}
L_{eff}= {2\pi \over 
{\sqrt {\Big(\Delta \cos^2\theta - U(1-\vec \beta \vec v_{f})\Big)^2+ \Delta^{2}\sin^{2} 2\theta}}}.
\label{l}
\end{equation}
Here $\Delta = {\delta m^{2}_\nu \over {2 |\vec p|}}$, $\delta m^{2}_\nu=m^{2}_{2}-m^{2}_{1}$ is the 
difference of the neutrino masses  squired, $\vec p$ is the neutrino momentum and $\theta$ is the vacuum 
mixing angle.

We can see that the neutrino oscillations characteristics, the probability $ P_{\nu_e \rightarrow \nu_\mu}(x)$, 
the  mixing angle $\theta_{eff}$ and 
the oscillation length  $ L_{eff} $, exhibit the dependence on the motion of matter. The resonance condition
\begin{equation} 
 {\delta m^{2}_\nu \over {2 |\vec p|}}\cos 2\theta= U(1-\vec \beta \vec v_{f}),
\label{res}
\end{equation}
at which the above probability has unit amplitude no matter how small the mixing angle $\theta$ is, 
also depends on the motion of matter. Therefore we conclude that the account for the 
relativistic motion of matter could provide the appearance of  the resonance in the neutrino oscillations in 
certain cases when for the given neutrino characteristics, 
$\delta m^{2}_\nu$, $|\vec p|$ and $\theta$,
and the invariant matter density at rest, $n_0$, the 
resonance is impossible. 

Let us rewrite  the resonance condition (\ref{res}) in terms of the invariant matter density:
\begin{equation} 
 {\delta m^{2}_\nu \over {2 |\vec p|}}\cos 2\theta= 
\sqrt {2} G_{F} {n_{0} \over {\sqrt {1-v^{2}_{f}}}}(1-\vec \beta \vec v_{f}).
\label{res0}
\end{equation}
The value of $n_{0}$ gives the matter density in the reference frame for which the total speed of matter is zero. 
That is why we can say that the effect of matter motion in the resonance condition is described by the factor
\begin{equation} 
{ 1-\vec \beta \vec v_{f} \over {\sqrt {1-v^{2}_{f}}}}.
\label{fac}
\end{equation}
If one estimates this factor for the ultra-relativistic neutrinos, $\beta \approx 1$, in the case when matter is moving
along the direction of the neutrino propagation also with high total speed then 
\begin{equation} 
{ 1-\vec \beta \vec v_{f} \over {\sqrt {1-v^{2}_{f}}}} \approx 
{\sqrt {1-v_{f}} \over \sqrt 2} \ll 1 .
\label{fac1}
\end{equation}
In the opposite case, when matter is moving against the direction of the neutrino propagation, one gets
\begin{equation} 
{ 1-\vec \beta \vec v_{f} \over {\sqrt {1-v^{2}_{f}}}} \approx {\sqrt 2 \over \sqrt {1-v_{f}} } \gg1 .
\label{fac2}
\end{equation}

From these estimations it follows that: 
1) the relativistic motion of matter along the neutrino propagation could 
provide the resonance increase of the oscillation probability if  the matter density $n_0$ is to high for the resonance 
appearance in non-moving matter,
2) the relativistic motion of matter in opposite direction to  the neutrino propagation could 
provide the resonance increase of the oscillation probability if  the matter density $n_0$ is to low for the resonance 
appearance in non-moving matter.

It worth to be noted that the similar analysis can be performed for any types of the neutrino 
flavour conversions and different matter composition and that effects of matter polarization 
can be also easily included \cite {GriLobStu02}. As it follows from the above 
consideration, there is 
an effective decrease (increase) of the matter effect in neutrino oscillations 
if matter 
is moving in the
direction that is parallel (opposite) to the neutrino propagation.
These phenomena have to be accounted in 
derivation of the resonance conditions for neutrino flavour 
(and also neutrino spin and spin flavour \cite {LobStuph01, DELSt01}) 
oscillations  
in the case when matter is 
moving with high speed.

\section*{Acknowledgements} 
We are thankful to E.Akhmedov, V.Berezinsky, S.Petcov, A.Smirnov and F.Vannucci 
for helpful discussions.


\begin{thebibliography}{99}
\bibitem {Wol78} L.Wolfenstein, Phys.Rev.D17 (1978) 2369.
\bibitem {MikSmi85} S.Mikheyev, A.Smirnov, Sov.J.Nucl.Phys.42 (1985) 913.
\bibitem {Wel82} H.Weldon, Phys.Rev.D26 (1982) 1394.
\bibitem {DOlNiePal89} J.D'Olivo, J.Nieves, P.Pal, Phys.Rev D40 (1989) 3679.

\bibitem{ELSt99} A.Egorov, A.Lobanov, A.Studenikin, in: New Worlds in
                Astroparticle Physics, ed. by A.Mourao, M.Pimento, P.Sa,
                World Scientific, Singapore, p.153, 1999; hep-ph/9902447.

\bibitem{ELStpl00} A.Egorov, A.Lobanov, A.Studenikin, Phys.Lett.B491 (2000)
                   137, hep-ph/9910476.


\bibitem{ELSt00} A.Egorov, A.Lobanov, A.Studenikin, in: Results and
                Perspectives in Particle Physics, ed. by M.Greco,
                Frascati Physics Series, v. 17, p. 117, 2000.
\bibitem {DvoStuYa01} M.Dvornikov, A.Studenikin, Phys.At.Nucl.64 (2001) 1.
\bibitem{DSthp0102099} M.Dvornikov, A.Studenikin, in:New Worlds in
                Astroparticle Physics, ed. by P.Sa,
                World Scientific, Singapore, 2001; in: Results and
                Perspectives in Particle Physics, ed. by M.Greco,
                Frascati Physics Series, 2001 (in press); hep-ph/0102099, hep-ph/0107109.
\bibitem {LobStuph01} A.Lobanov, A.Studenikin, Phys.Let.B515 (2001) 94; hep-ph/0106101.
\bibitem{DELSt01} M.Dvornikov, A.Egorov, A.Lobanov, A.Studenikin, in:
A.Studenikin (Ed.), Particle Physics at the Start of the New Millennium, 
World Scientific, Singapore, 2001, p. 178; hep-ph/0103015.
\bibitem {Pal92} P.Pal, Int.J.Mod.Phys.A7 (1992) 5387.
\bibitem{NSSV97} H.Nunokawa, V.Semikoz, A.Smirnov, J.W.F.Valle, Nucl.Phys.B501
(1997) 17.

\bibitem{BGN99} S.Bergmann, Y.Grossman, E.Nardi, Phys.Rev.D60 (1999) 093008.

\bibitem{GriLobStu02} A.Grigoriev, A.Lobanov, A.Studenikin, to be published.

\end{thebibliography}
\end{document}